\documentclass[journal=jpcbfk,manuscript=article,layout=traditional]{achemso}
\usepackage[version=3]{mhchem} 
\usepackage[T1]{fontenc}      
\usepackage{graphicx}
\usepackage{subfigure}
\usepackage{multirow}
\usepackage{hhline}
\usepackage{chemarr}
\usepackage{amssymb}
\usepackage{color}
\usepackage{xspace}
\usepackage{chngcntr}
\counterwithout{table}{section}
\usepackage{textcomp}
\usepackage{gensymb}
\usepackage{xr} 
\usepackage{pdfpages}
\usepackage{booktabs}
\usepackage{multirow}
\usepackage[table,xcdraw]{xcolor}
\usepackage{colortbl} 
\usepackage{bm}
\usepackage{float}
\usepackage[normalem]{ulem}

\DeclareRobustCommand{\del}[1]{\textcolor{red}{\sout{#1}}}

\usepackage{float}
\usepackage{graphicx} 
\usepackage{hyperref}
\usepackage{cleveref}
\usepackage{booktabs} 
\usepackage{graphicx} 

\usepackage{soul}
\usepackage{xcolor}
\usepackage{caption}
\usepackage{float}
\usepackage{float}
\usepackage{siunitx}

\usepackage{tikz}
\usetikzlibrary{positioning, arrows.meta}

\usepackage{marginnote}
\usepackage{enumitem}

\author{Narjes Ansari}\affiliation{Qubit Pharmaceuticals, Advanced Research Department, 75014 Paris, France}\email{narjesa@qubit-pharmaceuticals.com}
\author{Florent Hédin}\affiliation{Qubit Pharmaceuticals, Engineering Department, 75014 Paris, France}
\author{Jérôme Hénin}\affiliation{Laboratoire de Biochimie Théorique, Université Paris Cité, CNRS, 75005, Paris, France}
\author{Jean-Philip Piquemal}\affiliation{Qubit Pharmaceuticals, Advanced Research Department, 75014 Paris, France}\altaffiliation{Laboratoire de Chimie Théorique, Sorbonne Université, UMR 7616 CNRS, 75005 Paris, France}\email{jean-philip.piquemal@sorbonne-universite.fr}
\author{Louis Lagardère}\affiliation{Qubit Pharmaceuticals, Advanced Research Department, 75014 Paris, France}\altaffiliation{Laboratoire de Chimie Théorique, Sorbonne Université, UMR 7616 CNRS, 75005 Paris, France}\email{louis.lagardere@sorbonne-universite.fr}

\title{$\text{LAO}{+}$: Focused, Multi-Replica Accelerated Orthogonal Sampling for Robust Binding Free Energy Predictions}

\begin{document}

\maketitle
\begin{abstract}
We introduce Lambda-ABF-OPES+ ($\text{LAO+}$), a multi-replica framework for absolute binding free energy (ABFE) calculations that accelerates both the alchemical transformation and the slow ``orthogonal'' motions coupled to it. At the core of $\text{LAO+}$ is a partitioned $\lambda$-dynamics scheme that concentrates sampling on the abrupt rehydration of the binding pocket, the main bottleneck of ligand decoupling, while Hamiltonian replica exchange relays configurations along the entire alchemical path. This drives a fourfold gain in convergence speed for the rate-limiting leg over our previous $\text{LAO}$ method, which was itself up to ninefold faster than Lambda-ABF, an overall gain of up to $\sim$36-fold. Applied to 28 protein--ligand complexes across five targets, $\text{LAO+}$ achieves a root-mean-square error of 0.83~kcal~mol$^{-1}$ and a correlation of 0.87 with experiment from only 63~ns of sampling per ligand. Moreover, dedicated reservoir replicas accelerate any slow degree of freedom that can be described by a collective variable. We demonstrate this by recovering hidden protein and ligand conformational states, for example in T4 lysozyme, BRD4 and JNK1. Together, these advances make accurate ABFE calculations practical for drug discovery.

\end{abstract}

\section{Introduction}

Drug discovery is fundamentally driven by the thermodynamics of molecular recognition. Whether a ligand acts as a simple competitive inhibitor or triggers a complex conformational shift that modulates cell-signaling pathways, the success of a therapeutic agent is ultimately governed by its binding affinity. Optimizing this affinity while balancing safety and biological efficacy remains a primary challenge, making the accurate \textit{in silico} prediction of binding free energies a cornerstone of modern Computer-Aided Drug Design (CADD) \cite{de2008computational}. To reach the precision required for hit identification and lead optimization, researchers rely on atomistic Molecular Dynamics (MD) simulations to describe the time-dependent evolution of complexes involving ligands and their biological receptors, which may include proteins, nucleic acids (DNA/RNA), or other macromolecular targets.

The predictive power of these simulations is governed by two interdependent pillars: the accuracy of the underlying force field (FF) and the efficiency of the configurational sampling. On the accuracy front, traditional additive force fields often fail to capture the inhomogeneous electronic environments. This has led to the adoption of next-generation models, including polarizable force fields like AMOEBA \cite{ponder_current_2010, shi_polarizable_2013}, which explicitly account for many-body induction, and Neural Network Force Fields (NNFFs) such as Fennix Bio-1~\cite{ple2025foundation}. While these models provide \textit{ab initio}-level precision, their increased physical detail intensifies the ``sampling problem''. Biological phenomena, such as protein loop rearrangements or the displacement of buried water molecules, occur on timescales ranging from microseconds to milliseconds, or even hours---durations that far exceed the practical reach of standard, unbiased MD trajectories~\cite{pan2017quantitative}. Because slow sampling limits all of these models alike, the framework presented here acts on the sampling rather than on the energy model and is independent of the force field.

To bridge this gap, the community has developed a diverse array of enhanced sampling strategies categorized by their underlying mechanisms. The first category includes Collective Variable (CV)-based methods, such as Umbrella Sampling (US)~\cite{torrie1977nonphysical}, Adaptive Biasing Force (ABF)~\cite{darve2001calculating,darve2008adaptive,comer2015adaptive}, MetaDynamics (MtD)~\cite{laio2002escaping,barducci2011metadynamics}, its recent evolution, On-the-fly Probability Enhanced Sampling (OPES)~\cite{Invernizzi2020,Invernizzi2022}, and Temperature-Accelerated Molecular Dynamics (TAMD)~\cite{maragliano2006temperature}, which apply external forces along specific reaction coordinates. These CVs can be either \textit{physical-based}, targeting spatial coordinates, or \textit{alchemical-based}, where it corresponds to a scaling, denoted as $\lambda$, of selected interactions between parts of the molecular system, effectively enabling their decoupling. Starting from a fully interacting Hamiltonian, standard protocols first scale down electrostatic interactions (electrostatic leg or ELE leg), and then van der Waals interactions (van der Waals leg or VDW leg) while also leveraging softcores in order to avoid the so called endpoint catastrophe and improve convergence\cite{beutler1994avoiding}. Simulations also require estimators such as  Free Energy Perturbation (FEP)~\cite{zwanzig1954high} or Thermodynamic Integration (TI)~\cite{straatsma1991multiconfiguration}. Traditional approaches exploit trajectories at different fixed $\lambda$ values but another approach, $\lambda$-dynamics \cite{kong1996lambda}, assigns a mass to the alchemical variable making it vary alongside the Cartesian coordinates.  Alchemical methods have gained widespread adoption in drug discovery due to their ability to provide a rigorous thermodynamic framework for calculating binding affinities. Despite these advancements, alchemical Absolute Binding Free Energy (ABFE) calculations~\cite{mobley2012perspective} are frequently hindered by ``orthogonal'' barriers---structural bottlenecks that do not lie directly along the $\lambda$-coordinate but are coupled to the alchemical state. A primary example is the reorganization of the hydration shell: as a ligand's volume is decoupled in the VDW leg, the binding pocket must undergo a transition from a dry to a solvated state~\cite{aldeghi2016accurate, bhati2024equilibrium}. This water-exchange process is often slow and characterized by discrete hydration states, leading to significant hysteresis and slow convergence in standard alchemical cycles~\cite{anand2026improving}. Even recent high-throughput schemes that minimize protein--ligand relative motion along the alchemical path, such as the Lucid double-decoupling method~\cite{bian2025formally}, identify water exchange into and out of the binding site as their main remaining limitation. Beyond solvation, the alchemical path is often obstructed by other slow conformational degrees of freedom. These include host-side bottlenecks such as protein side-chain rotamer transitions~\cite{baumann2021challenges, osato2026automated} and loop movements, as well as ligand-side bottlenecks such as the interconversion between distinct bound conformations, macrocyclic strain adaptations, or the trapping of internal ligand torsions~\cite{sasmal2020sampling}.

Another category of techniques addresses the sampling problem through parallelism by simulating multiple realizations of the system simultaneously~\cite{Swendsen1986,Zuckerman2017,Husic2018}. These approaches can be broadly divided into \textit{multi-walker} and \textit{multi-replica} frameworks based on their exchange mechanism. In multi-walker strategies, such as Multiple-Walkers Metadynamics (MW-MetaD)~\cite{laio2002escaping,minoukadeh2010potential}, identical realizations of the system cooperate by contributing to a shared, evolving bias potential to map the free energy surface more rapidly. Conversely, multi-replica frameworks, such as Hamiltonian Replica Exchange (HREX)~\cite{sugita1999replica,Mitsutake2004}, utilize replicas with varying internal parameters---or Hamiltonians---and periodically attempt to exchange their coordinates based on a physical acceptance criterion. A promising way forward is to combine the power of these parallel simulations with CV-based enhanced sampling to efficiently cross both enthalpic and entropic barriers. A prominent example of an alchemical-based approach is FEP+~\cite{wang2015accurate_v2}, which utilizes a multi-replica framework (FEP/REST2~\cite{wang2011rest2_v2}) where the molecular Hamiltonian is scaled to determine precise binding free energies. Similarly, Jiang et al.~\cite{jiang2018reduced} introduced a reduced FEP/H-REMD scheme that combines a standard 1D alchemical thermodynamic axis with Hamiltonian-tempering boosted replicas. By attaching accelerated replicas exclusively to the alchemical end-states, this approach facilitates the relay of configurations across the fixed-$\lambda$ windows to overcome ``Hamiltonian lagging'' caused by slow structural reorganizations. Expanding on this philosophy of combining multi-replica architectures with adaptive biasing, the recently introduced OneOPES framework~\cite{rizzi2023oneopes} powerfully integrates these two domains. 

Analogously, the multi-walker Lambda-ABF method~\cite{lagardere2024labf} represents a powerful alternative within this hybrid family. In our previous study, we demonstrated that layering OPES-Explore ($\text{OPES}_e$)~\cite{Invernizzi2022} on top of the Lambda-ABF scheme—a combined approach we refer to as L-ABF-OPES ($\text{LAO}$)~\cite{ansari2025lambda}—achieved up to a 9-fold acceleration in convergence compared to standard L-ABF method. Our results indicated that $\text{OPES}_e$ accelerates the $\lambda$ dynamics and significantly aids in the exploration of orthogonal degrees of freedom, such as the critical process of pocket hydration within the binding pocket~\cite{ansari2025lambda,blazhynska2025water}. Consistent with the physical nature of alchemical transformations, our previous findings highlighted a distinct transition in the hydration state of the binding site. Specifically, a clear discontinuity in the number of binding-site water molecules occurs near $\lambda \approx 0.2$ during the VDW decoupling leg as the ligand's volume is phased out and water molecules rush to occupy the vacated space. While the exploratory nature of $\text{OPES}_e$ facilitates transitions between these hydration states, this universal ``water jump'' represents a severe orthogonal kinetic barrier. Consequently, a single walker performing stochastic continuous $\lambda$-dynamics across the entire unpartitioned range exhibits significant ``Hamiltonian lagging''—spending disproportionate computational time in fully hydrated or completely dry end-states while rarely executing the critical back-and-forth transitions across the actual barrier region. This inherent complexity causes the VDW leg to converge significantly slower than the electrostatic leg, where water rearrangement is less dramatic.

To better manage these abrupt physical transitions and eliminate deep-seated orthogonal traps, we present Lambda-ABF-OPES+ ($\text{LAO+}$). Philosophically, $\text{LAO+}$ is designed as a physical interpolation between the continuous exploration of $\lambda$-dynamics and the intrinsically guaranteed sampling (per $\lambda$ window) of fixed-$\lambda$ windowing frameworks. This framework builds directly on the continuous sampling power of our previous methodology but replaces the global multi-walker architecture with an interconnected, multi-replica framework based on HREX. This advanced architectural shift yields three synergistic advantages: First, $\text{LAO+}$ divides the global alchemical schedule into smaller, and purposefully non-uniform lambda intervals. By focusing multiple replicas exclusively within the $\lambda$ regions governing the dry-to-wet phase transition, we achieve a far more optimal allocation of computational resources, forcing the system to repeatedly sample critical hydration/dehydration states rather than over-sampling trivial end-states. Second, while such partitioning inherently restricts the continuous global motion of any single walker, this spatial limitation is seamlessly overcome by the active HREX layer. By periodically attempting coordinate swaps between neighboring replicas based on the Metropolis-Hastings criterion, configurations (such as a perfectly hydrated pocket conformation generated in a transition window) are relayed across the discrete windows to maintain fluid, global phase-space mixing and prevent Hamiltonian lagging. Third, this multi-replica setup allows for the definition of specialized, state-based thermodynamic ``reservoirs'' running localized orthogonal biases. Crucially, as established by our detailed validation, while the HREX layer acts as a minor efficiency booster in standard smoothly converging landscapes, it becomes absolutely essential when deep orthogonal barriers are present. In this configuration, targeted replicas act as enhanced sampling engines for slow degrees of freedom—such as discrete water-displacement states or host side-chain rotamers—and actively inject these rare-event configurations back into the physical alchemical ensemble via rigorous metropolized coordinate exchanges, fundamentally curing local kinetic trapping.

To demonstrate the robustness of this architecture across diverse physical bottlenecks, we first validate the integrated framework on ABFE calculations across five distinct ligand--protein datasets encompassing the Trypsin--Benzamidine complex, 11 BRD4 inhibitors, 5 CHK1 inhibitors, 6 PWWP1 inhibitors, and 5 P38 kinase inhibitors.  In these core validation tracks, by effectively refreshing the structural ensemble and bridging hydration discontinuities, $\text{LAO+}$ successfully resolves kinetic traps that pose significant sampling challenges and require prohibitive computational times in standard MD and 1D alchemical methods. Consequently, this integrated framework achieves a four-fold acceleration in convergence of the rate-limiting complex VDW leg compared to our previous hybrid approach. Furthermore, $\text{LAO+}$ significantly improves computational efficiency by reducing the required number of parallel trajectories; whereas the original $\text{LAO}$ utilized four walkers for both complex and solvent phases across the ELE and VDW legs, $\text{LAO+}$ achieves precise thermodynamic convergence using only three replicas for the complex electrostatic and solvent phases.

Beyond standard alchemical scaling, we further test the capability of our specialized reservoir architecture to handle severe host and ligand sampling bottlenecks that typically stall 1D alchemical coordinates. On the protein side, we target the rotameric fluctuations of Val111 in the L99A mutant of T4 Lysozyme and Asn140 in BRD4, both of which represent significant hurdles that frequently trap standard alchemical simulations in local minima. On the ligand side, we target the slow $180^{\circ}$ ring-flips commonly observed in JNK1 kinase inhibitors. In these final cases representing both protein and ligand challenges, the use of specialized reservoirs running a localized OPES bias enables the efficient exploration and subsequent injection of these hidden orthogonal states back into the physical ensemble via HREX. This targeted approach ensures a complete and thermodynamically accurate sampling of discrete conformational transitions, allowing the simulation to reach the true global free energy minimum without getting trapped in unrepresentative local states.

\section{Methods: The $\text{LAO+}$ Framework}

Below we will describe briefly the mathematical foundation behind the HREX algorithm as it was implemented in Tinker-HP.

\paragraph{Setup}
Consider two replicas $a$ and $b$ with $i = \{a,b\}$ that explore the phase space under different conditions (modified Hamiltonian, different target temperatures or target pressures, etc.); the HREX algorithm will attempt periodic coordinate swaps between them using the following equations to validate or reject the swap.

Let:
$H_i(\mathbf{q_i})$ be the Hamiltonian of replica $i$ evaluated at coordinates $\mathbf{q}$;
$\beta_i = 1 / (k_B T_i)$ be the inverse temperature of replica $i$;
$P_i$ be the target (NPT) or instantaneous (NVT) pressure of replica $i$;
$V_i$ be the target (NVT) or instantaneous (NPT) volume of replica $i$.

\paragraph{NVT Contribution}
For a pure Hamiltonian (lambda) exchange the energy difference is:

$$
\delta E^{a\leftrightarrow b} = \bigl[H_b(\mathbf{q}_a) + H_a(\mathbf{q}_b)\bigr] - \bigl[H_a(\mathbf{q}_a) + H_b(\mathbf{q}_b)\bigr]
$$

In the general Hamiltonian + temperature (lambda + $T$) case, the reduced energy difference is:

$$
\Delta E_\text{NVT}^{a\leftrightarrow b}
  = \bigl[\beta_b\,H_b(\mathbf{q}_a) - \beta_a\,H_a(\mathbf{q}_a)\bigr]
  + \bigl[\beta_a\,H_a(\mathbf{q}_b) - \beta_b\,H_b(\mathbf{q}_b)\bigr]
$$

(when $\beta_a = \beta_b = \beta$ this reduces to the pure-lambda case $\Delta E_\text{NVT}^{a\leftrightarrow b} = \beta\,\delta E^{a\leftrightarrow b}$).

\paragraph{NPT Contribution}

For an isobaric ensemble the reduced Hamiltonian includes a $PV$ work term. The additional contribution to the exchange criterion is therefore:

$$
\Delta E_\text{NPT}^{a\leftrightarrow b}
  = (\beta_a\,P_a - \beta_b\,P_b)\,(V_b - V_a)
$$

\paragraph{Total Acceptance Criterion}

The total acceptance criterion therefore always contains the $\Delta E^{a\leftrightarrow b}_\text{NVT}$ term, to which is added the isobaric contribution if the simulation runs in the NPT ensemble:
$$
\Delta E^{a\leftrightarrow b} = \Delta E^{a\leftrightarrow b}_\text{NVT} + \Delta E^{a\leftrightarrow b}_\text{NPT}
$$

Global ergodicity and detailed balance are maintained through the following Metropolis-Hastings inspired acceptance criterion:

$$
P^{a\leftrightarrow b}_\text{acc} = \min\bigl(1,e^{-\Delta E^{a\leftrightarrow b}}\bigr)
$$

\subsection{The Reservoir Concept for Orthogonal Acceleration}
The central innovation of $\text{LAO+}$ is the ability to define specialized \text{reservoirs} that act as engines for accelerating orthogonal degrees of freedom (DOF). These reservoirs are essentially replicas where the Hamiltonian or the ensemble is modified to bypass kinetic traps that would otherwise hinder the convergence of the physical alchemical path. This architecture enables two primary modes of acceleration:
\begin{itemize}
\item \textbf{Biased Reservoirs:} A reservoir replica can be configured to apply an additional OPES bias along a specific physical coordinate (e.g., a dihedral angle, a distance, or water coordination). This allows the system to cross high-energy barriers in the orthogonal space and ``inject'' these rare-event configurations back into the physical $\lambda$-states via HREX swaps.
\item \textbf{Thermal Reservoirs:} Alternatively, a reservoir can be defined within a Parallel Tempering (Replica Exchange) framework. By maintaining one or more replicas at a higher temperature ($T_{high}$), the system can more easily navigate rugged potential energy surfaces. These high-temperature structures are then propagated back to the lower-temperature physical replicas, ensuring that the alchemical path samples from the true global distribution.
\end{itemize}
Thermal reservoirs are available in the implementation but were not used in the present work; all reservoirs below are biased reservoirs.

\subsection{Non-linear Topology and Custom Exchange Networks}
To maximize sampling efficiency, $\text{LAO+}$ moves beyond standard nearest-neighbor exchange protocols. In simple simulation scenarios, the framework utilizes a linear chain of replicas (e.g., Rep 0, 1, and 2), where coordinate swaps are restricted to immediate neighbors. While this is sufficient for basic systems, complex scenarios involving high kinetic barriers require a more flexible architecture. By implementing a user-defined connectivity matrix, the framework establishes a non-linear topology that allows for a more complex exchange graph.

\begin{figure}[htbp]
    \centering
    \includegraphics[width=0.7\textwidth]{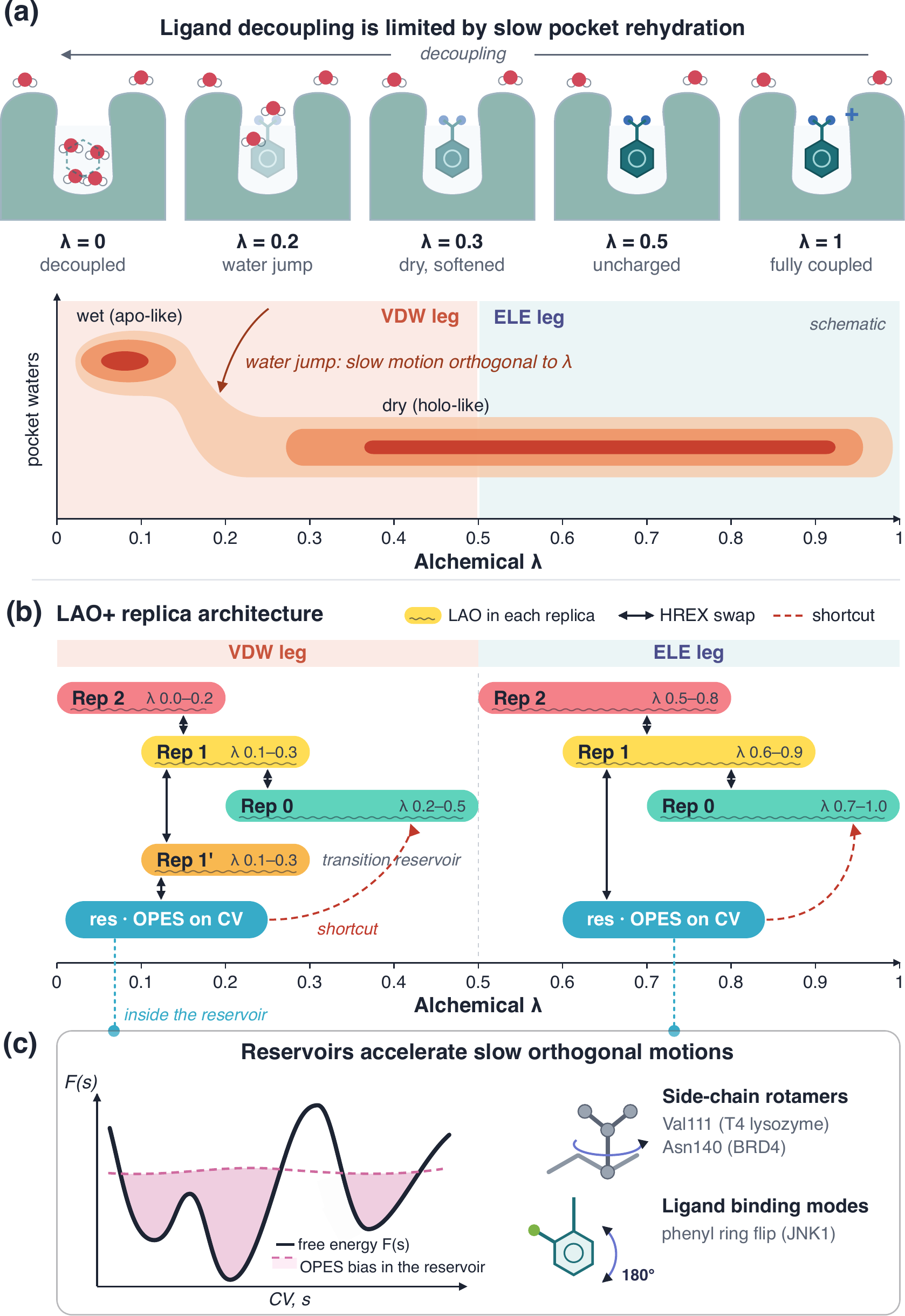}
\caption{\textbf{Overview of the $\text{LAO+}$ framework.}
(a) Alchemical decoupling: charges are removed along the ELE leg ($0.5 \le \lambda \le 1$) and steric volume along the VDW leg ($0 \le \lambda \le 0.5$). Near $\lambda \approx 0.2$ the pocket abruptly rehydrates (``water jump''), a slow motion orthogonal to $\lambda$ that limits convergence of the VDW leg (data in Fig.~\ref{fig:lao_solvent_sampling}).
(b) Replica architecture on the same $\lambda$ axis. Each replica runs Lambda-ABF-OPES (LAO) within its own window, and replicas exchange configurations via HREX. VDW windows are concentrated on the water jump, with Rep~2 and Rep~0 acting as wet and dry reservoirs, Rep~1 as a bridge, and Rep~1$'$ as a transition reservoir; ELE windows are uniform. Conformational reservoirs (res) can be attached to either leg and feed rare configurations into the chain through HREX and shortcuts.
(c) Within a reservoir, OPES biases a slow orthogonal CV $s$, filling the basins so that barriers are crossed freely, e.g., side-chain rotamers (Val111, T4 lysozyme; Asn140, BRD4) or ligand ring flips (JNK1).}
\label{fig:LAOP_overview}
\end{figure}

As illustrated in Fig.~\ref{fig:LAOP_overview}b, this connectivity network can be extended based on the specific complexity of the system. This allows for auxiliary replicas, such as Rep 1', to function alongside the primary alchemical path. Rep 1' can be configured identically to Rep 1 in terms of its alchemical state but serves as a dedicated engine for sampling acceleration. Because Rep 1' is permitted to exchange with both its direct alchemical neighbor (Rep 1) and other defined states, it creates a high-connectivity hub.

The framework offers various possibilities for placing reservoirs (res) depending on the structural requirements of the target. These reservoirs provide ``shortcuts'' that facilitate the rapid propagation of rare conformations across the entire alchemical schedule, helping the system overcome significant structural bottlenecks such as ligand-ring flipping or protein side-chain reorientation. As shown in Fig.~\ref{fig:LAOP_overview}b, a reservoir can be attached to either the VDW or the ELE leg, where it exchanges with the alchemical replicas through HREX swaps and dedicated shortcuts. While multiple reservoirs can be implemented to address highly complex protein environments, the current study achieved convergence using a single reservoir strategy to manage the conformational transitions effectively. This architecture ensures that all states can refresh their structural ensembles through direct exchange with global sampling engines.

\subsubsection{Accelerating Pocket Hydration Bottlenecks in ABFE}
\begin{figure}[H]
    \centering
    \includegraphics[width=\textwidth]{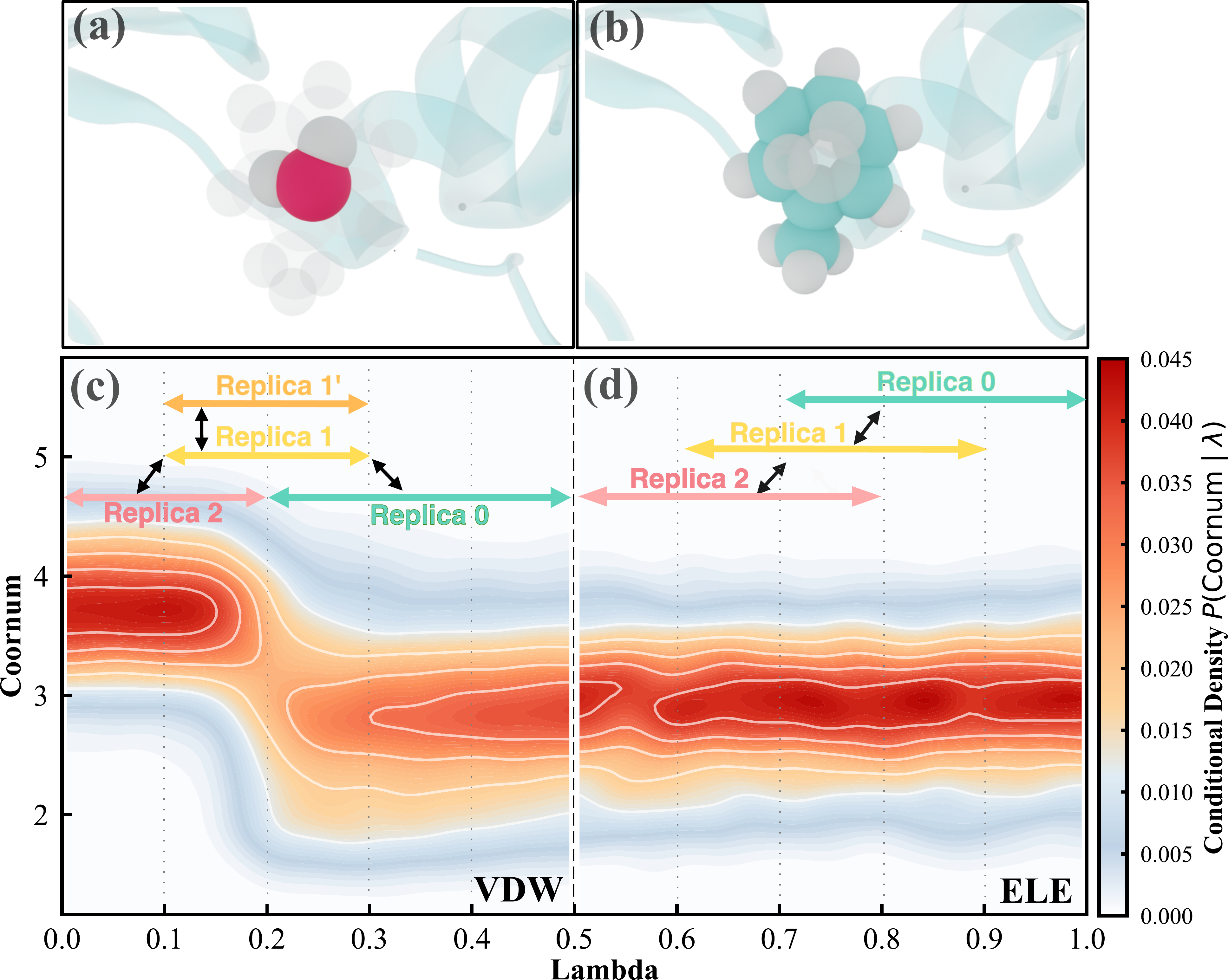}
   \caption{\textbf{Schematic representation of the $\text{LAO+}$ partitioning framework for accelerating pocket re-hydration bottlenecks.} 
    (a) Representative snapshot of the fully solvated apo-like state, where the physical ligand volume is decoupled ($\lambda = 0$) and a structured water molecule occupies the binding site. 
    (b) Representative snapshot of the dry holo-like state, featuring the fully coupled ligand packing within the binding pocket. 
    (c) Two-dimensional (2D) column-normalized conditional probability density landscape, $P(\text{Coornum} \mid \lambda)$, for the VDW leg. A sharp hydration phase discontinuity (the 'water jump') is cleanly resolved near $\lambda \approx 0.2$, marking the severe orthogonal kinetic barrier between the solvated apo-like and dry holo-like basins. The non-uniform replica partitioning strategy concentrates overlapping resources (Replica 1 and $1'$) directly over this transition region to accelerate configuration relays via HREX. 
    (d) Conditional density landscape for the ELE leg, exhibiting a smoothly continuous coordination profile. The corresponding uniform, equidistant replica allocation reflects the absence of abrupt solvation phase transitions during electrostatic decoupling.}
    \label{fig:lao_solvent_sampling}
\end{figure}

A primary challenge in ABFE calculations is the reorganization of the hydration shell. As shown in Fig.~\ref{fig:lao_solvent_sampling}, the hydration landscape of the binding pocket varies significantly across the alchemical trajectory. During the ELE decoupling leg ($0.5 < \lambda < 1.0$), the water coordination remains essentially constant; because the ligand's steric volume is already fully present, the solvent environment is stable and lacks discrete transitions. Conversely, the VDW decoupling leg ($0 < \lambda < 0.5$) is characterized by a sharp transition in the number of water molecules near $\lambda \approx 0.2$. This rehydration is slow and orthogonal to $\lambda$, which makes it the main convergence bottleneck of the VDW leg (Fig.~\ref{fig:LAOP_overview}a).

Despite the improvements introduced in our previous work, where we demonstrated that OPES$_{e}$ could accelerate the sampling of orthogonal degrees of freedom, our results indicated that the convergence of the VDW leg remained significantly slower than that of the ELE leg. 

To resolve the problem—where simulations remain kinetically trapped in configurations inconsistent with the current alchemical state —the $\text{LAO+}$ framework utilizes a strategic, multi-replica architecture based on partitioned and overlapping $\lambda$ ranges. This approach is designed to relay configurations across the entire alchemical axis, ensuring that accelerated states from thermodynamic reservoirs can refresh the physical ensemble. As shown in the upper portions of Fig.~\ref{fig:lao_solvent_sampling}{c--d}, the replica scheduling differs between the two alchemical legs based on the underlying solvent landscape. The VDW leg employs a non-uniform reservoir scheduling to address the water-sampling challenge shown in Fig. \ref{fig:lao_solvent_sampling}c. This landscape reveals a sharp hydration discontinuity at $\lambda \approx 0.2$, where the system transitions from a ``wet'' apo-like state (see Fig.~\ref{fig:lao_solvent_sampling}a) to a ``dry'' holo-like state (see Fig.~\ref{fig:lao_solvent_sampling}b). To bridge this bottleneck, the alchemical schedule is partitioned into specific ranges with significant overlap between neighbors to facilitate high-frequency HREX swaps. Each range is sampled by its own LAO replica (Lambda-ABF-OPES within the window), as depicted in Fig.~\ref{fig:LAOP_overview}b.

In this configuration, Replica 2 ($0.0 \le \lambda \le 0.2$) is used to explore the sampling of the wet-like ensemble, which also acts as a reservoir for the hydrated apo-like state. Similarly, Replica 0 ($0.2 \le \lambda \le 0.5$) explores the dry-like ensemble and serves as a reservoir for the dehydrated holo-like state. Replica 1 ($0.1 \le \lambda \le 0.3$) is strategically positioned to overlap

 with both, effectively serving as a bridge between these two distinct reservoirs. By linking Replica 1 to these two stable states, the framework accelerates the exploration of solvent reorganization by ``injecting'' correctly-sampled configurations from both the wet and dry basins into the transition region. 

Furthermore, to resolve the high energetic cost of solvent reorganization in the complex phase, we introduce a specialized extra replica, Replica 1' ($0.1 \le \lambda \le 0.3$), which acts as an additional transition reservoir to boost sampling directly at the structural bottleneck. In contrast, the ELE leg shown in Fig. \ref{fig:lao_solvent_sampling}d utilizes a uniform range scheduling. 

Because this leg exhibits a continuous coordination density without a discrete structural transition or ``gap'' in the solvent environment, equidistant partitioning is sufficient to maintain global ergodicity. The lambda ranges for Replica 2 ($0.5 \le \lambda \le 0.8$), Replica 1 ($0.6 \le \lambda \le 0.9$), and Replica 0 ($0.7 \le \lambda \le 1.0$) are spaced uniformly but still retain substantial overlap to ensure high exchange acceptance ratios. By utilizing these partitioned ranges and the HREX cascade, $\text{LAO+}$ bypasses high-energy barriers by importing sampled configurations from the stable reservoirs into the rest of the ensemble, removing the need for additional exploratory biases on solvent degrees of freedom.

\subsubsection{Accelerating Protein and Ligand Conformational Sampling}

Beyond hydration bottlenecks, the $\text{LAO+}$ architecture provides the capability to define specialized thermodynamic reservoirs to accelerate other slow degrees of freedom, including protein side-chain rotamer flips and ligand binding mode inter-conversions. These transitions are notoriously difficult to capture using standard alchemical paths, as the system often remains kinetically trapped in configurations inconsistent with the thermodynamic end-states. As illustrated in Fig.~\ref{fig:LAOP_overview}b--c, a replica can be configured as a conformational reservoir (res) where an OPES bias is applied directly to a specific physical CV, such as a dihedral angle or a distance. By applying these boosting potentials to flatten the free energy surface along targeted coordinates, the system can escape kinetic traps that would otherwise lead to ``Hamiltonian lagging'' and hysteresis. Depending on the specific sampling challenge, these reservoirs are strategically linked to the alchemical axis via HREX to facilitate coordinate swaps. For instance, to investigate protein-side transitions, we utilize extra replicas to explore the rotameric fluctuations of Val111 in T4 Lysozyme and the Asn140 side-chain flips in BRD4. This reservoir architecture also provides a robust solution to the ``frozen rotamer'' problem in ligands. As noted in previous studies, ligands often bind in multiple distinct modes separated by significant kinetic barriers that prevent spontaneous inter-conversion during standard MD timescales. While methods like the BLUES~\cite{sasmal2020sampling} framework utilize hybrid MD/NCMC approaches to facilitate these transitions, $\text{LAO+}$ offers an alternative via its multi-replica exchange cascade. Following the principle of alchemical region selection—where only the ``floppy'' or moving part of the ligand is perturbed to reduce work variance—our reservoir defines this specific region as a localized alchemical zone. By enabling the exchange of these rare protein and ligand conformations across the $\lambda$ landscape, $\text{LAO+}$ ensures that correctly sampled configurations are ``injected'' into the physical ensemble. This approach ensures the calculated binding affinity reflects the true thermodynamic equilibrium rather than a kinetically trapped state, providing a high-precision solution for complex binding environments.

\section{Simulation Details}

All simulations were performed using the massively parallel molecular dynamics package \texttt{Tinker-HP} (Version 1.2)~\cite{tinker-hp,adjoua_tinker-hp_2021}, which incorporates the \texttt{Colvars} library~\cite{fiorin2013using} for the implementation of $\lambda$-dynamics and the \textit{LAO+} multi-replica framework. The AMOEBA force field~\cite{ponder_current_2010,wu2012automation,shi_polarizable_2013,AMOEBA-nucleic} was employed to describe the protein, water, and ions; the $\text{LAO+}$ framework itself acts on the sampling and does not depend on the choice of energy model. Ligand parameterization was conducted using the \texttt{Poltype 2} package~\cite{poltype2}. The simulated systems consisted of approximately 31,000 to 58,000 atoms, depending on the specific protein target and ligand size.

For all systems, a cubic water box was utilized, ensuring a minimum distance of 12~\AA\ between the ligand and the periodic boundaries. The Benzamidine--Trypsin system was neutralized without the addition of extra salt, following the protocol established in Ref.~\citenum{ansari2022water}. The L99A T4 Lysozyme system was neutralized with 50~mM physiological NaCl according to Ref.~\citenum{sasmal2020sampling}, while all other complexes were neutralized with a 0.15~M physiological NaCl concentration.

The equilibration protocol followed a multi-step procedure. Initial energy minimization was performed using the \texttt{Tinker-HP} \texttt{minimize} utility. Subsequently, the systems underwent a two-step heating process from 200~K to 298~K in the $NVT$ ensemble. In the first 4~ns stage, harmonic restraints were applied to all protein--ligand atoms and crystal water molecules. In the second 4~ns stage, restraints were restricted to heavy backbone atoms using a force constant of 10~kcal/mol/\AA$^2$. These initial stages utilized the RESPA integrator with a 2.0~fs time step. 

In the third equilibration step, we employed the three-level multiple timestep BAOAB-RESPA1 integrator~\cite{lagardere_pushing_2019} with an outer timestep of 10~fs (intermediate: 3.33~fs; inner: 1~fs) in the $NVT$ ensemble for 5~ns, with 1~kcal/mol/\AA$^2$ restraints applied only to C$\alpha$ atoms. This was followed by a 5~ns $NPT$ equilibration at 1~atm. Production runs, ranging from 20 to 100~ns depending on the ligand, were conducted in the $NPT$ ensemble using the BAOAB-RESPA1 Langevin integrator with a 10~fs outer time step. 

Temperature was controlled via the Bussi thermostat~\cite{bussi2007canonical} and pressure was regulated with the Berendsen barostat~\cite{berendsen1984molecular}. Van der Waals interactions were treated with a 12~\AA\ cutoff, while electrostatic interactions were computed using Particle Mesh Ewald (PME)~\cite{essmann1995smooth} with a 7~\AA\ real-space cutoff. Induced dipoles were solved to a convergence tolerance of $1 \times 10^{-5}$~Debye using a Preconditioned Conjugate Gradient (PCG) solver~\cite{lagardere2015scalable}.

For the $\text{LAO+}$ simulations, the system setups were identical to those used in plain MD, with the exception of the integrator, which employed the BAOAB-RESPA algorithm with a 3~fs time step. For the ABFE calculations, the VDW leg of the complex was simulated using four replicas (replica 0, replica 1, replica 1', and replica 2; see Fig.~\ref{fig:lao_solvent_sampling}c) for a total production time of 9~ns per replica. In contrast, the ELE leg of the complex employed three replicas (replica 0, replica 1, and replica 2; see Fig.~\ref{fig:lao_solvent_sampling}d), each running for 3~ns. For the solvent phase, both the ELE and VDW legs were sampled using three replicas (replica 0, replica 1, and replica 2), with a simulation length of 3~ns per replica. Consequently, the total cumulative simulation time allocated per ligand was 63~ns per independent repeat (45~ns for the complex phase and 18~ns for the solvent phase, summed over all replicas); three independent repeats were performed for each ligand.

\begin{figure}[p]
	\centering
    \includegraphics[width=1\textwidth]{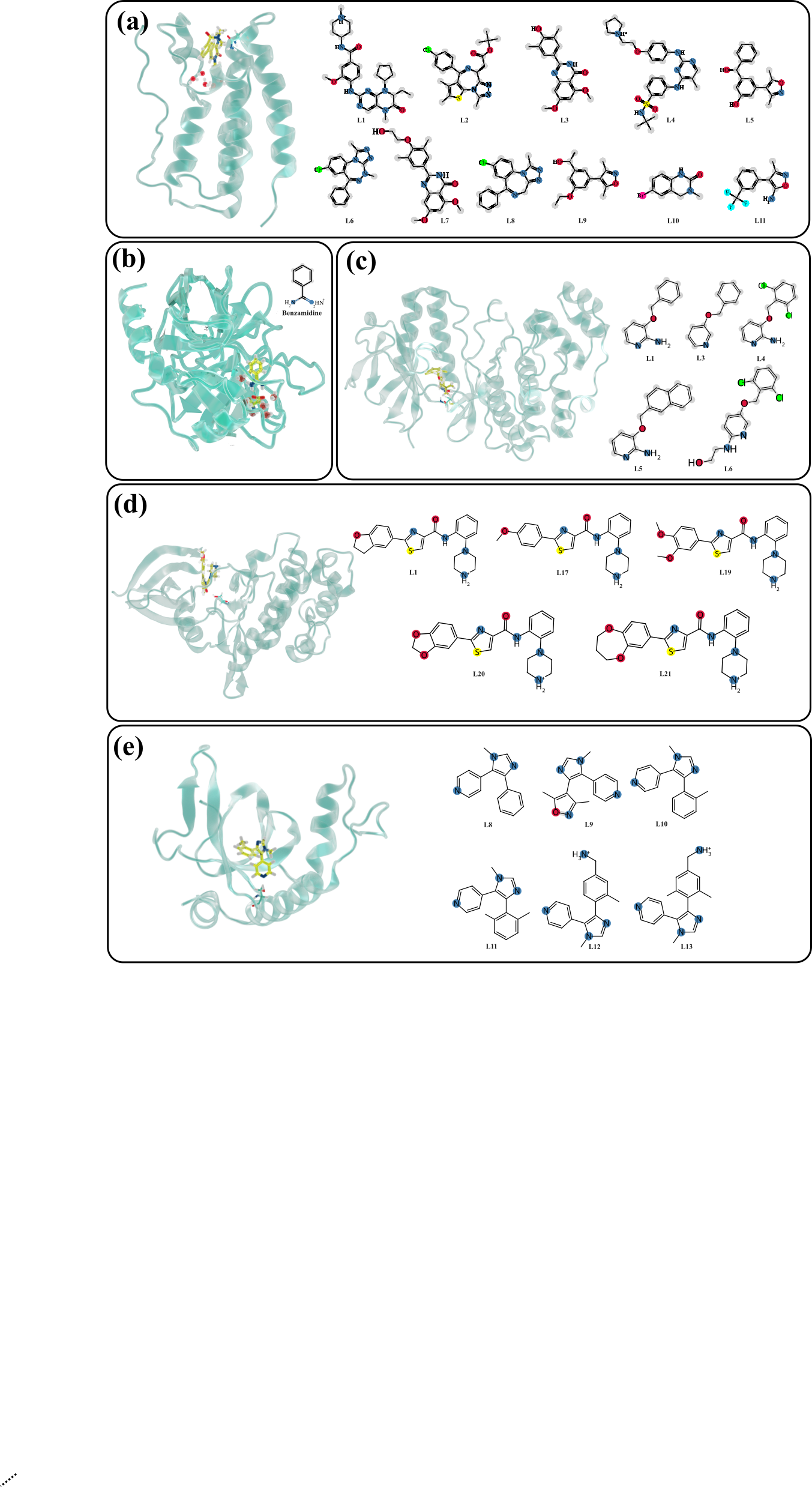}
\end{figure}
\begin{figure}[t]
	\caption{\textbf{Structural overview of the protein--ligand benchmark sets}.Cartoon representation of the binding pockets and corresponding 2D chemical structures of the ligands for: (a) BRD4 in complex with 11 drug-like inhibitors; (b) the Trypsin--Benzamidine complex; (c) P38 kinase with 5 ligands; (d) CHK1 with 5 inhibitors; and (e) PWWP1 with 6 ligands. In each protein panel, the binding site is shown with representative ligands.}
	\label{fig:all_systems}
\end{figure}

\section{Results}

\subsection{Benchmark systems}

To demonstrate the robustness of our approach, we selected a diverse set of targets that exemplify these challenges, including the Trypsin–Benzamidine complex and 11 BRD4 inhibitors, six PWWP1 inhibitors, five P38 kinase inhibitors, and five CHK1 inhibitors, see Fig.~\ref{fig:all_systems}.

Our analysis begins with the Trypsin and BRD4 systems, which were previously studied using the original L-ABF-OPES methodology. By re-evaluating these targets with the $\text{LAO+}$ framework, we can directly quantify the impact of the multi-replica architecture on convergence rates and sampling efficiency. Following this comparison, we present the results for the remaining systems to demonstrate the general applicability of the method across a broader range of chemical space and structural challenges.

\subsection{Water-Mediated Interactions in BRD4 and Trypsin}

To evaluate the impact of Hamiltonian replica exchange on the convergence of the L-ABF-OPES method, we recalculated the binding affinities for a benchmark set previously studied using the original approach. This set includes 11 drug-like ligands in complex with BRD4 and the Benzamidine--Trypsin complex (see Fig.~\ref{fig:all_systems}a-b). To ensure a direct comparison and isolate the effect of $\text{LAO+}$ on sampling efficiency, we maintained all simulation inputs, force field parameters, and binding pocket definitions from the previous study~\cite{ansari2025lambda}. For the BRD4 systems, we specifically employed the same restraint protocol and corrections to handle the rotameric states of Asn140. This involved a mild positional restraint on the residue to prevent artifacts during alchemical transitions and the application of a $-RT\ln(2)$ correction to the final $\Delta G$ to account for apo-state degeneracy; the reader is referred to the original paper~\cite{ansari2025lambda} for further technical details regarding this Asn140 rotamer treatment. This focus on identical parameterization ensures that any observed improvements in convergence can be attributed solely to the multi-replica framework rather than differences in the underlying physical model.

\begin{figure}[htbp]
    \centering
    \includegraphics[width=0.9\textwidth]{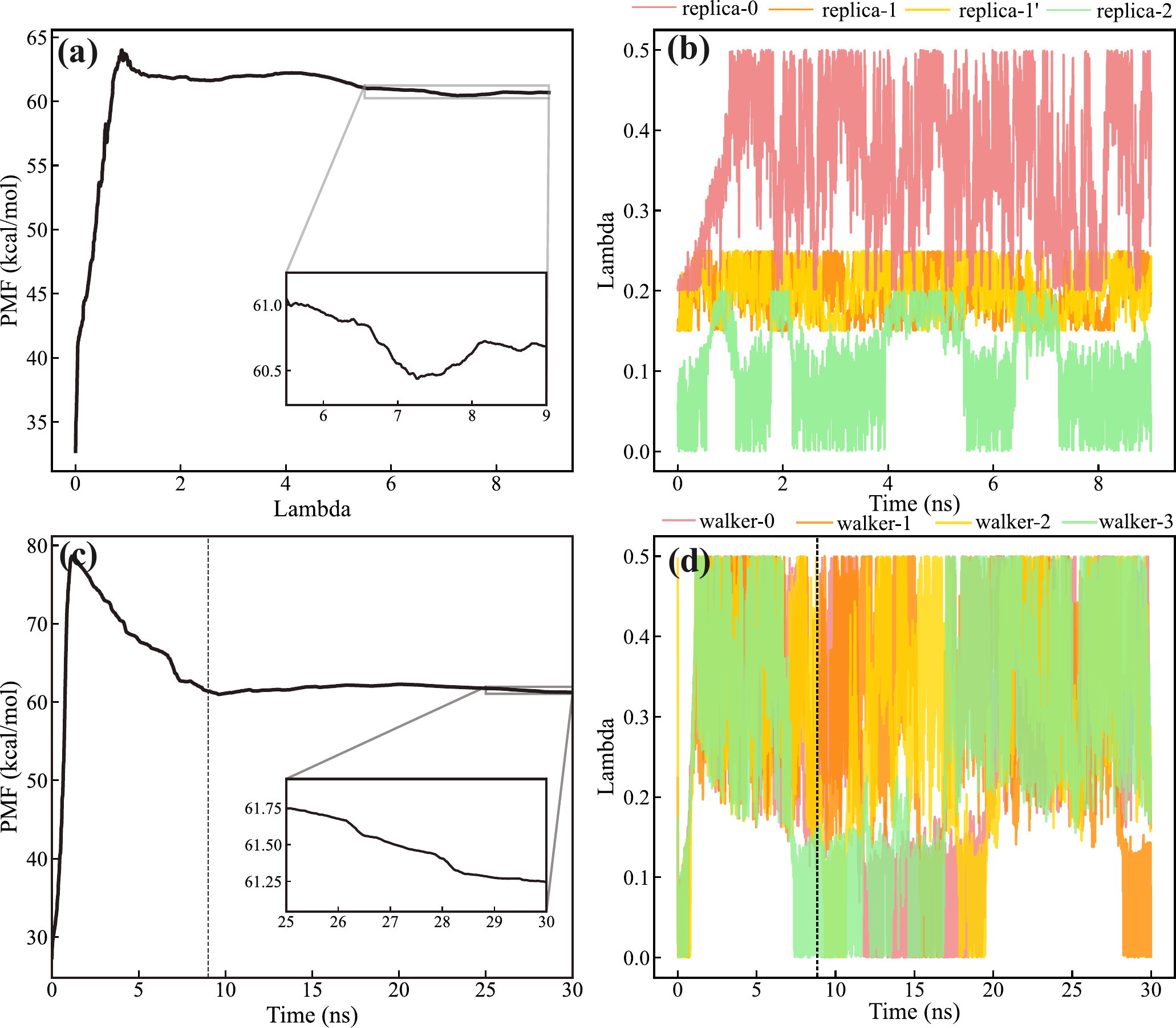}
    \caption{\textbf{Comparison of convergence and sampling efficiency between $\text{LAO+}$ and $\text{LAO}$ for BRD4 Ligand 1.} 
    Top panels display results for the $\text{LAO+}$ framework: \textbf{(a)} PMF convergence for the VDW leg, with a stabilization after approximately 5.5~ns, and \textbf{(b)} $\lambda$ trajectory fluctuations across the four replicas (0, 1, $1'$, and 2). 
    Bottom panels display the corresponding results for the standard $\text{LAO}$ method: \textbf{(c)} PMF convergence and \textbf{(d)} $\lambda$ fluctuations for the four walkers. 
    The vertical dashed line indicates the 9~ns mark, representing the total simulation time for $\text{LAO+}$, while the $\text{LAO}$ simulation was extended to 30~ns for comparison.
    }
    \label{ref:LAO_vs_LAOP_BRD4_LIG1_PMF_LAMBDA}
\end{figure}

To assess the efficiency of the multi-replica architecture, we present a showcase of the PMF convergence and $\lambda$ sampling dynamics for the VDW leg of Ligand 1 in the BRD4 set, as illustrated in Fig.~\ref{ref:LAO_vs_LAOP_BRD4_LIG1_PMF_LAMBDA}. As discussed previously, the VDW leg represents the primary sampling challenge due to the hydration discontinuity and the transition between apo-like and holo-like states during re-hydration. As shown in Fig.~\ref{ref:LAO_vs_LAOP_BRD4_LIG1_PMF_LAMBDA}a, the $\text{LAO+}$ framework achieves rapid convergence within the first 9~ns of simulation time. The inset in the VDW leg PMF plot confirms that the free energy estimate stabilizes significantly after approximately 5.5~ns.

In contrast, the standard $\text{LAO}$ method (Fig.~\ref{ref:LAO_vs_LAOP_BRD4_LIG1_PMF_LAMBDA}c) exhibits a much slower convergence profile, requiring nearly 25~ns to reach a comparable level of stability.

To quantify these timelines rigorously, the convergence time was defined—similar to our previous work~\cite{ansari2025lambda}—as the point from which all subsequent data points remain within a specified tolerance of the final mean. Here a strict tolerance of 0.3~kcal/mol was used in this analysis. 

This discrepancy in convergence speed is directly attributable to the enhanced sampling of the alchemical space, particularly within the critical transition region. In the $\text{LAO+}$ framework (Fig.~\ref{ref:LAO_vs_LAOP_BRD4_LIG1_PMF_LAMBDA}b), the $\lambda$ fluctuations demonstrate that replicas undergo multiple round-trips across the alchemical axis within the 9~ns window, maintaining high sampling density around $\lambda \approx 0.2$. This specific region represents the structural transition state where water is displaced or recruited. Conversely, the standard $\text{LAO}$ simulations (Fig.~\ref{ref:LAO_vs_LAOP_BRD4_LIG1_PMF_LAMBDA}d) show far more sluggish dynamics; at the 9~ns mark (indicated by the dashed line), the walkers have barely completed a single transition, leaving the transition region at $\lambda \approx 0.2$ almost entirely unsampled during this period. This increased frequency of back-and-forth transitions in $\text{LAO+}$ facilitates the rapid ``injection'' of diverse configurations from the thermodynamic reservoirs into the bottleneck region, which is the primary driver for the accelerated convergence observed. The robust connectivity of the $\text{LAO+}$ replica exchange cascade is further evidenced by the acceptance statistics for this system. 

For all systems, coordinate exchanges were attempted every 1000 steps, corresponding to an exchange frequency of 3~ps given the 3~fs time step. For the ELE leg, a standard nearest-neighbor exchange protocol was employed among the three replicas. At each exchange interval, swaps were attempted alternately between pairs, alternating between the Replica 0 $\leftrightarrow$ 1 and Replica 1 $\leftrightarrow$ 2 pairs, with all pairs receiving equal weight. Under this uniform scheduling, we observed an excellent acceptance ratio of approximately 50\% for each exchanging pair, indicating highly fluid mixing across the electrostatic landscape.

Conversely, the VDW leg required a non-uniform, weighted random exchange schedule. Because a single replica cannot participate in multiple exchange attempts simultaneously, the presence of the parallel branch (Replica 1') prevents a simple sequential nearest-neighbor arrangement. The physical state to the first transition replica (Replica 0 $\leftrightarrow$ 1) was assigned a weight of four; the transition-to-transition swap (Replica 1 $\leftrightarrow$ 1') was given a weight of one; and the bridging pathways leading to the second transition state (Replica 1 $\leftrightarrow$ 2 and Replica 1' $\leftrightarrow$ 2) were each assigned a weight of two. 

Under this weighted scheme, we observed highly successful configuration swaps throughout the cascade. For a representative target, the final exchange acceptance ratio between Replica 0 $\leftrightarrow$ 1 was 5.33\%, while high-frequency connectivity was maintained between the transition reservoirs (Replica 1 $\leftrightarrow$ $1'$) at approximately 35\%. Furthermore, the bridging swaps between the specialized transition replicas and the fully hydrated reservoir (Replica 1 $\leftrightarrow$ 2 and $1'$ $\leftrightarrow$ 2) showed consistent acceptance rates of 9.25\% and 9.08\%, respectively. 

To isolate the individual impacts of lambda partitioning and active replica swapping, an ablation study was performed on Ligand 1 and Ligand 5 of the BRD4 dataset, as well as on alanine dipeptide in water, by disabling HREX swaps and conducting simulations using only the partitioned lambda windows (detailed in Supplementary Section S5).

Table~\ref{tab:LAO_BRD4} summarizes the calculated binding free energies ($\Delta G_{bind}$) obtained with the $\text{LAO+}$ framework, compared against results from the previous $\text{LAO}$ study~\cite{ansari2025lambda} and experimental values. Additionally, the table reports the convergence times for both the $\text{LAO}$ and $\text{LAO+}$ methods to highlight the impact of the replica exchange framework on computational efficiency. The correlation between the $\text{LAO}$ and $\text{LAO+}$ results is illustrated in Fig. S1 
Using the $\text{LAO+}$ method, we achieved an RMSE of 0.9~kcal/mol and a Pearson’s correlation coefficient ($r$) of 0.84, compared to 1.1~kcal/mol and 0.81 for the original $\text{LAO}$ method. While $\text{LAO+}$ provides a slightly improved RMSE, the most significant advantage is the acceleration in sampling. For the VDW leg of the complex phase, convergence was achieved approximately four times faster than with the $\text{LAO}$ method. Although we did not observe a significant reduction in convergence time for the ELE leg, it is noteworthy that $\text{LAO+}$ achieved these results using only three replicas, whereas the original $\text{LAO}$ protocol required four. This demonstrates that the Hamiltonian exchange between replicas not only improves the reliability of the solvent-filled and ligand-occupied states but also significantly reduces the total computational cost required to reach a converged binding free energy.

\begin{table}[htbp]
\centering
\scriptsize
\setlength{\tabcolsep}{3pt} 
\resizebox{\textwidth}{!}{
\begin{tabular}{lcccccccccccc}
\toprule
& & & & & \multicolumn{8}{c}{\textbf{Convergence Time ($t_{\text{conv}}$, ns)}} \\ \cmidrule(lr){6-13} 
& & & & & \multicolumn{4}{c}{\textbf{Complex}} & \multicolumn{4}{c}{\textbf{Solvent}} \\ \cmidrule(lr){6-9} \cmidrule(lr){10-13}
& & & & & \multicolumn{2}{c}{\textbf{ELE}} & \multicolumn{2}{c}{\textbf{VDW}} & \multicolumn{2}{c}{\textbf{ELE}} & \multicolumn{2}{c}{\textbf{VDW}} \\ \cmidrule(lr){6-7} \cmidrule(lr){8-9} \cmidrule(lr){10-11} \cmidrule(lr){12-13}
\multirow{-4}{*}{\textbf{Comp.}} & \multirow{-4}{*}{$\mathbf{\Delta G_{exp}}$} & \multirow{-4}{*}{$\mathbf{\Delta G_{LAO+}}$} & \multirow{-4}{*}{$\mathbf{\Delta G_{LAO}}$} & \multirow{-4}{*}{\textbf{PDB}} & \textbf{LAO+} & \textbf{LAO} & \textbf{LAO+} & \textbf{LAO} & \textbf{LAO+} & \textbf{LAO} & \textbf{LAO+} & \textbf{LAO} \\ \midrule
L1  & -9.8 $\pm$ 0.1 & -9.89 $\pm$ 0.78 & -10.56 $\pm$ 0.46 & 4OGI  & 1.6 & 0.9 & 5.6 & 22.5 & 1.7 & 1.9 & 1.9 & 3.5 \\
L2  & -9.6 $\pm$ 0.1 & -8.54 $\pm$ 0.65 & -8.26 $\pm$ 0.22  & 3MXF  & 1.5 & 2.2 & 6.0 & 19.2 & 1.3 & 0.9 & 2.2 & 2.2 \\
L3  & -9.0 $\pm$ 0.1 & -8.40 $\pm$ 0.40 & -9.19 $\pm$ 0.30  & 4MR3  & 1.9 & 2.6 & 5.2 & 17.8 & 1.1 & 0.4 & 1.7 & 2.4 \\
L4  & -8.9 $\pm$ 0.1 & -9.29 $\pm$ 0.59 & -10.13 $\pm$ 0.15 & 4OGJ  & 2.0 & 2.6 & 5.9 & 24.8 & 1.7 & 2.7 & 2.0 & 3.5 \\
L5  & -8.8 $\pm$ 0.1 & -8.42 $\pm$ 0.44 & -8.27 $\pm$ 0.52  & 4J0R  & 1.1 & 2.1 & 5.5 & 15.9 & 0.8 & 0.9 & 1.3 & 2.7 \\
L6  & -8.2 $\pm$ 0.1 & -7.78 $\pm$ 0.66 & -6.75 $\pm$ 0.30  & 3U5L  & 0.8 & 0.6 & 6.2 & 20.2 & 0.4 & 0.5 & 2.4 & 3.5 \\
L7  & -7.8 $\pm$ 0.1 & -8.30 $\pm$ 0.24 & -6.42 $\pm$ 0.65  & 4MR4  & 1.5 & 0.9 & 5.5 & 20.0 & 1.1 & 1.4 & 2.2 & 3.4 \\
L8  & -7.4 $\pm$ 0.1 & -5.55 $\pm$ 0.58 & -5.46 $\pm$ 0.53  & 3U5J  & 0.7 & 0.7 & 5.8 & 20.7 & 0.5 & 0.4 & 0.8 & 2.5 \\
L9  & -7.3 $\pm$ 0.1 & -6.85 $\pm$ 0.54 & -7.57 $\pm$ 0.12  & 3SVG  & 0.9 & 1.5 & 6.3 & 17.4 & 0.4 & 0.5 & 1.7 & 2.7 \\
L10 & -6.3 $\pm$ 0.0 & -6.89 $\pm$ 0.33 & -7.05 $\pm$ 0.33  & 4HBV  & 0.9 & 1.8 & 3.7 & 14.8 & 3.7 & 0.2 & 1.2 & 1.6 \\
L11 & -5.6           & -6.03 $\pm$ 0.68 & -4.98 $\pm$ 0.58  & Model & 0.2 & 0.2 & 5.7 & 10.7 & 0.9 & 0.2 & 1.3 & 1.9 \\ 
Benzamidine-Trypsin & -6.36 & -6.60 $\pm$ 0.43 & -6.22 $\pm$ 0.65 & 3ATL & 1.7 & 1.0 & 5.0 & 17.0 & 1.4 & 0.34 & 1.7 & 3.5 \\ \bottomrule
\end{tabular}
}
\caption{\textbf{Thermodynamic binding free energies and convergence times for BRD4 and Trypsin datasets.} Reported binding free energies ($\Delta G$) and convergence times ($ t_{\text{conv}} $) represent the mean values over  and associated standard errors calculated across three independent replicate simulations.} 
\label{tab:LAO_BRD4}
\end{table}

\subsection{ABFE Validation on CHK1, P38, and PWWP1 Datasets}

\begin{figure}[H]
    \centering
    \includegraphics[width=0.6\textwidth]{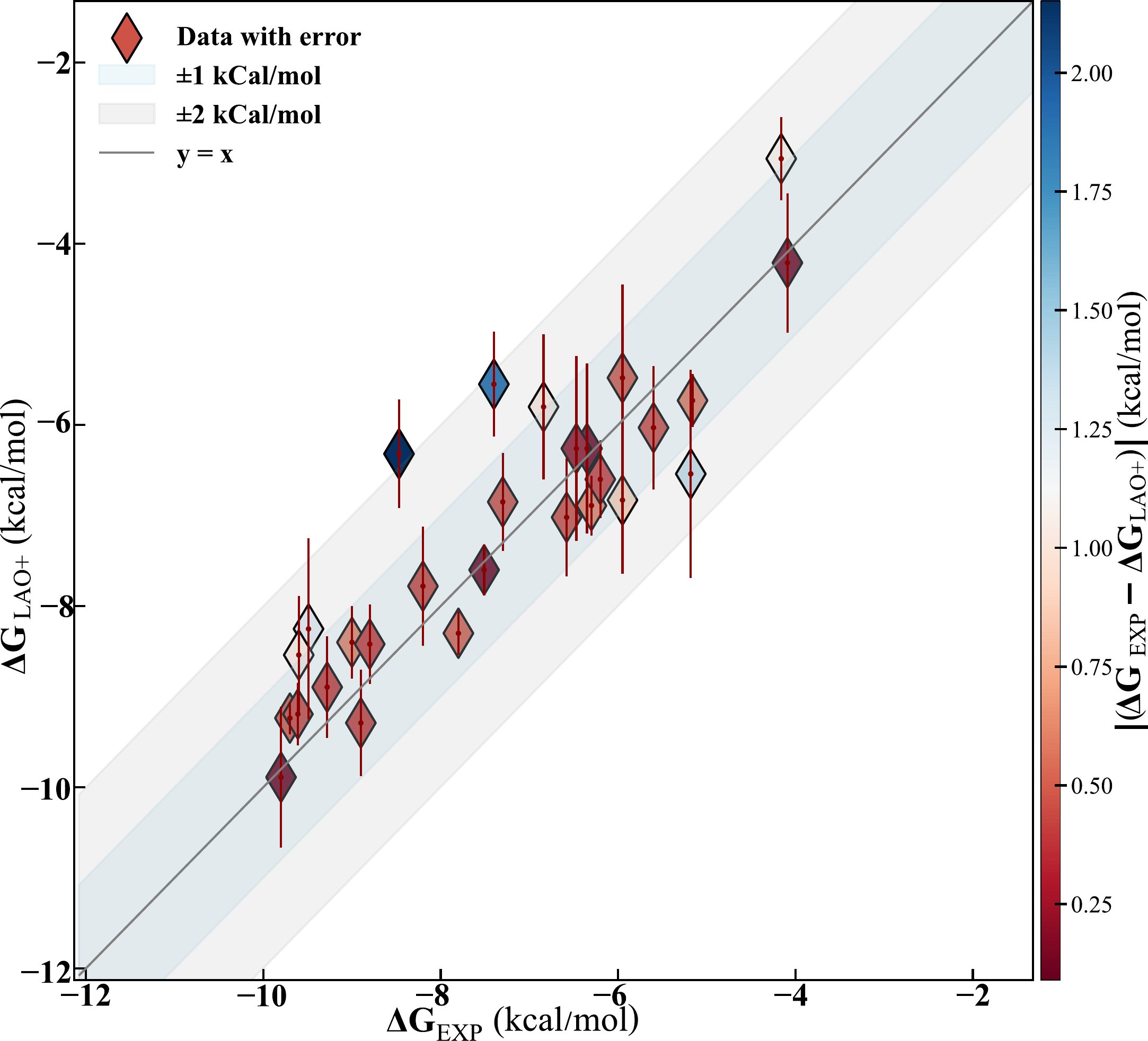}
    \caption{\textbf{Assessment of binding free energy accuracy and convergence consistency for all studied systems (BRD4, Trypsin, CHK1, PWWP1 and P38).} 
    (a) Correlation between experimental ($\Delta G_{exp}$) and calculated binding free energies using the $\text{LAO+}$ framework ($\Delta G_{LAO+}$). The dashed line represents the $y=x$ ideal correlation, with shaded regions indicating deviations within $\pm1$~kcal/mol and $\pm2$~kcal/mol. 
    Data points are shown as diamonds with error bars; points are colored by the absolute difference $|\Delta G_{EXP} - \Delta G_{LAO+}|$ as indicated by the color scale.
Calculated $\Delta G$ values and their associated errors represent the mean and standard error derived from three independent repeats for each ligand.}
\label{fig:EXP_vs_LAOP_All}
\end{figure}

To further evaluate the robustness of the $\text{LAO+}$ framework, we applied the method to three additional targets: CHK1 in complex with 5 ligands, P38 in complex with 5 ligands, and PWWP1 in complex with 6 ligands (Fig.~\ref{fig:all_systems}c--e). Consistent with the results observed for BRD4 and Trypsin, the $\text{LAO+}$ method provided high-quality predictions of binding free energies ($\Delta G_{bind}$) using the same protocol of 63~ns of total sampling per ligand and repeat. Figure~\ref{fig:EXP_vs_LAOP_All} illustrates the correlation between experimental and simulated values across all 28 complexes studied in this work. Overall, the framework achieved an RMSE of 0.83~kcal/mol and a Pearson’s correlation coefficient ($r$) of 0.87, demonstrating the accuracy and broad applicability of the method. 

In the case of P38, several ligands induce an apo-to-holo conformational change (Supplementary Section S4). Because the short $\text{LAO+}$ trajectories cannot sample this reorganization reproducibly for every ligand and replica, we treated all P38 complexes consistently by applying a mild C$\alpha$ restraint (1~kcal~mol$^{-1}$~\AA$^{-2}$) that keeps the protein in its holo conformation. The calculated values therefore omit the apo-to-holo reorganization free energy, which, to a first approximation, is common to all ligands of the series. We accounted for it with a single constant shift, $\Delta G_{\text{shift}} = \langle\Delta G_{\text{exp}}\rangle - \langle\Delta G_{\text{calc}}\rangle = 5$~kcal~mol$^{-1}$, averaged over the P38 series and added to all P38 values~\cite{chen2023enhancing,ansari2025targeting}. This shift fixes the absolute scale but leaves the relative affinities unchanged, so the P38 set validates relative rather than absolute binding free energies.

\subsection{Valine 111 Reorientation During 3-Iodotoluene Binding}

The L99A mutant of T4 lysozyme provides a simple example of induced conformational change upon ligand binding, centered within its buried hydrophobic cavity. While the system is structurally straightforward, certain ligands such as 3-iodotoluene induce slow side-chain rearrangements that must be accurately modeled to determine the binding free energy. Specifically, the Valine 111 (Val111) side chain must rotate to create sufficient space for the ligand. We utilized the LAO+ framework to evaluate the efficiency of the Hamiltonian replica exchange architecture in capturing this slow conformational mode.

\begin{figure}[H]
    \centering
    \includegraphics[width=0.7\textwidth]{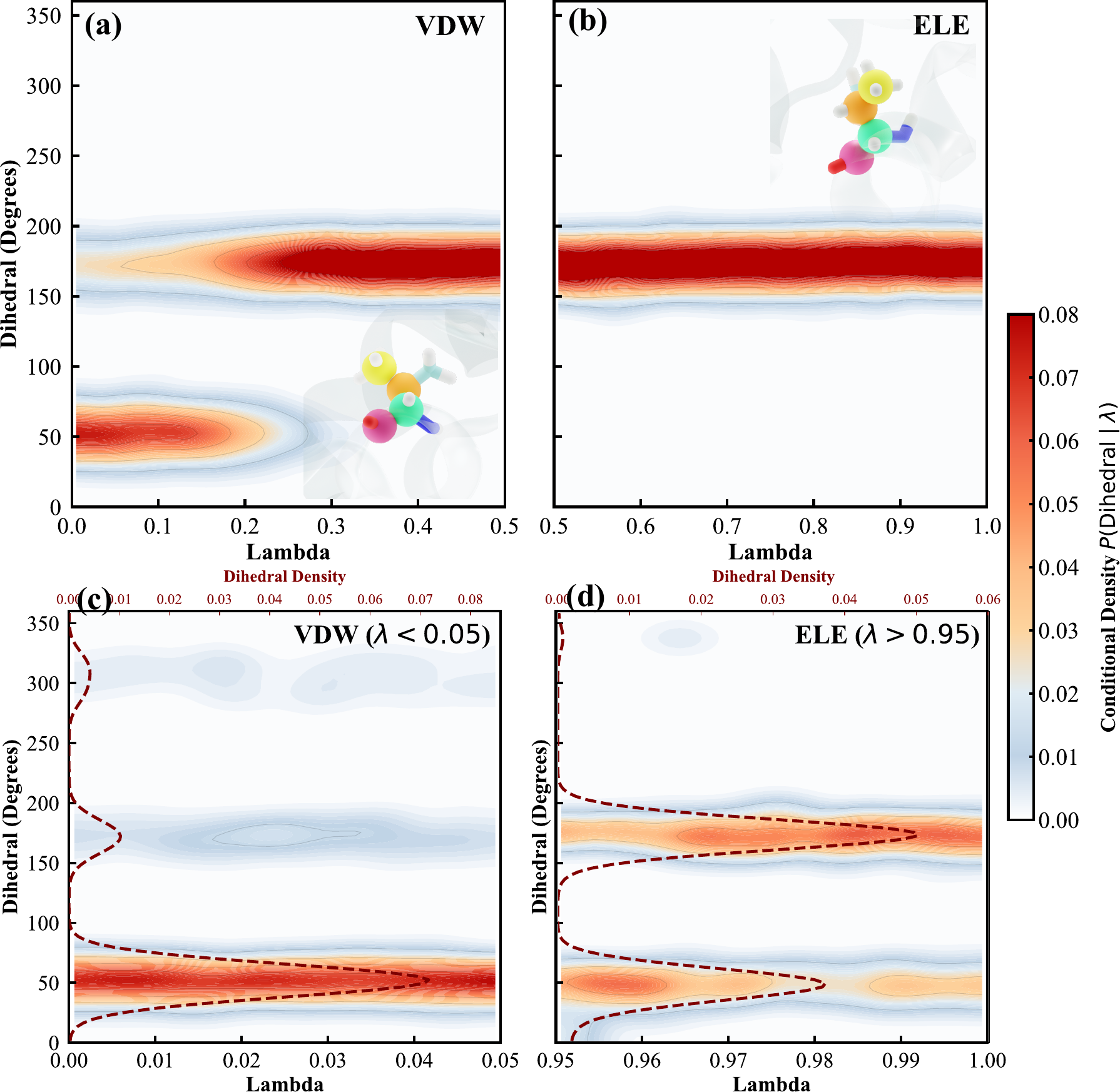}
    \caption{\textbf{Conditional side-chain rotamer landscapes of Val111 in the T4 lysozyme L99A system using the $\text{LAO+}$ framework.} 
    Two-dimensional (2D) column-normalized conditional probability density landscapes, $P(\text{Dihedral} \mid \lambda)$, are tracked across the full alchemical coordinate for the standard $\text{LAO+}$ framework in (a) the VDW region and (b) the ELE region, simulated without a conformational reservoir. 
    Panels (c) and (d) display the corresponding landscapes utilizing $\text{LAO+}$ with the reservoir architecture. These panels map out the endpoint-localized conditional landscapes alongside matching 1D probability density projections (dashed maroon lines) at the VDW ($\lambda < 0.05$) and ELE ($\lambda > 0.95$) milestones, respectively, achieved by engaging the OPES-biased conformational reservoir. }
    \label{fig:dihedral_sampling_reservoir}
\end{figure}

We executed the $\text{LAO+}$ scheme using the $\lambda$-scheduling previously discussed. Throughout both the VDW and the ELE legs, we monitored the Val111 dihedral angle(see Fig.\ref{fig:dihedral_sampling_reservoir} a-b for the definition of dihedral). The resulting density plots of the dihedral versus $\lambda$ are shown in Fig.~\ref{fig:dihedral_sampling_reservoir}. In the ELE regime without a specialized reservoir (see Fig.~\ref{fig:dihedral_sampling_reservoir}b), the holo-like state is characterized by a high-probability rotamer with a dihedral angle centered around $175^{\circ}$, which remains the dominant conformation. In contrast, the VDW leg—which transitions through apo-like states—reveals a more complex landscape (see Fig.~\ref{fig:dihedral_sampling_reservoir}a). Specifically for $\lambda < 0.2$, the system explores two distinct rotameric basins at approximately $55^{\circ}$ and $175^{\circ}$, suggesting that in the absence of full ligand sterics, both orientations are thermally accessible. 

To validate these baseline observations, we performed an independent OPES-enhanced simulation of the true apo state. As shown in the 1D Free Energy Surface (FES) profile (Supplementary Fig.~S4), the true apo landscape has its global minimum at $55^{\circ}$ ($g^{+}$), with the $305^{\circ}$ ($g^{-}$) and $175^{\circ}$ ($trans$) basins lying 2.2 and 3.3~kcal~mol$^{-1}$ higher, corresponding to Boltzmann populations of approximately 97\%, 2.4\% and 0.4\% at 298~K. The barrier tops lie about 7.0~kcal~mol$^{-1}$ ($g^{+}\leftrightarrow trans$, near $117^{\circ}$), 8.8~kcal~mol$^{-1}$ ($trans \leftrightarrow g^{-}$, near $235^{\circ}$) and 5.5~kcal~mol$^{-1}$ ($g^{-}\leftrightarrow g^{+}$, near $355^{\circ}$) above the global minimum. The $305^{\circ}$ basin is thus the second most populated apo rotamer. This $305^{\circ}$ conformation is completely omitted during standard alchemical sampling due to the high kinetic barriers, highlighting a significant sampling deficiency that typically compromises standard 1D alchemical protocols.

To overcome this bottleneck, we exploited our specialized reservoir architecture by incorporating an additional, localized OPES-biased replica into the framework. For the VDW leg, the reservoir replica was targeted to map the region around $\lambda \in [0.0, 0.2]$, while for the ELE leg, it covered $\lambda \in [0.5, 0.8]$. Within these reservoir replicas, a relatively low biasing potential ($W \approx 3$~kcal/mol) was applied on the Val111 dihedral coordinate using the OPES method (see SI for more details). Critically, trajectories from these biased states are completely excluded from the final Potential of Mean Force (PMF) and free energy reconstructions; instead, they function strictly as a conformational engine designed to actively explore and inject hidden rare-event configurations back into the physical alchemical pathway via HREX.

The success of this reservoir injection mechanism is clearly illustrated in the 2D dihedral density distributions as a function of $\lambda$ (Fig.~\ref{fig:dihedral_sampling_reservoir}c--d). With the addition of the OPES-biased reservoirs, the active exploration of the orthogonal phase space allows the system to bridge these high barriers and unearth all three physical basins. Near the apo-like endpoint ($\lambda < 0.05$), the 1D probability density profile superimposed on the contour plot demonstrates that the system now resolves three distinct peaks matching the true apo landscape: a dominant peak at $55^{\circ}$, a secondary peak at $175^{\circ}$, and the previously missing minor population at $305^{\circ}$. Conversely, as the system tracks toward the fully coupled holo-like endpoint ($\lambda > 0.95$), the scaling interactions cleanly shift the structural ensemble; while a population is still visibly sampled near $50^{\circ}$, the landscape exhibits a clear thermodynamic dominance centered around the $175^{\circ}$ rotameric state, with the secondary basins being smoothly suppressed as the physical environment is restored.

This improvement in sampling directly impacts the accuracy of the binding free energy estimates. Without the reservoir architecture, the unrefined simulation yielded a $\Delta G_{\text{bind}}$ of $-5.89 \pm 0.49$~kcal/mol.

 Upon engaging the specialized reservoirs to achieve a complete and thermodynamically accurate sampling of all three rotameric states, the converged value refined to $-5.29 \pm 0.39$~kcal/mol. While an explicit experimental value for this specific ligand-mutant pair is unavailable in the literature, our reservoir-refined result lies within the range reported by previous computational studies, which report values between $-4.8$ and $-5.7$ ~kcal/mol~\cite{baumann2021challenges}. The 0.6~kcal~mol$^{-1}$ shift is comparable to the combined statistical uncertainty; the principal result is that the reservoir recovers the $g^{-}$ population that standard alchemical sampling never visits.

\subsection{Asn140 Reorientation During BRD4 Binding}

\begin{figure}[htbp]
    \centering
    \includegraphics[width=\textwidth]{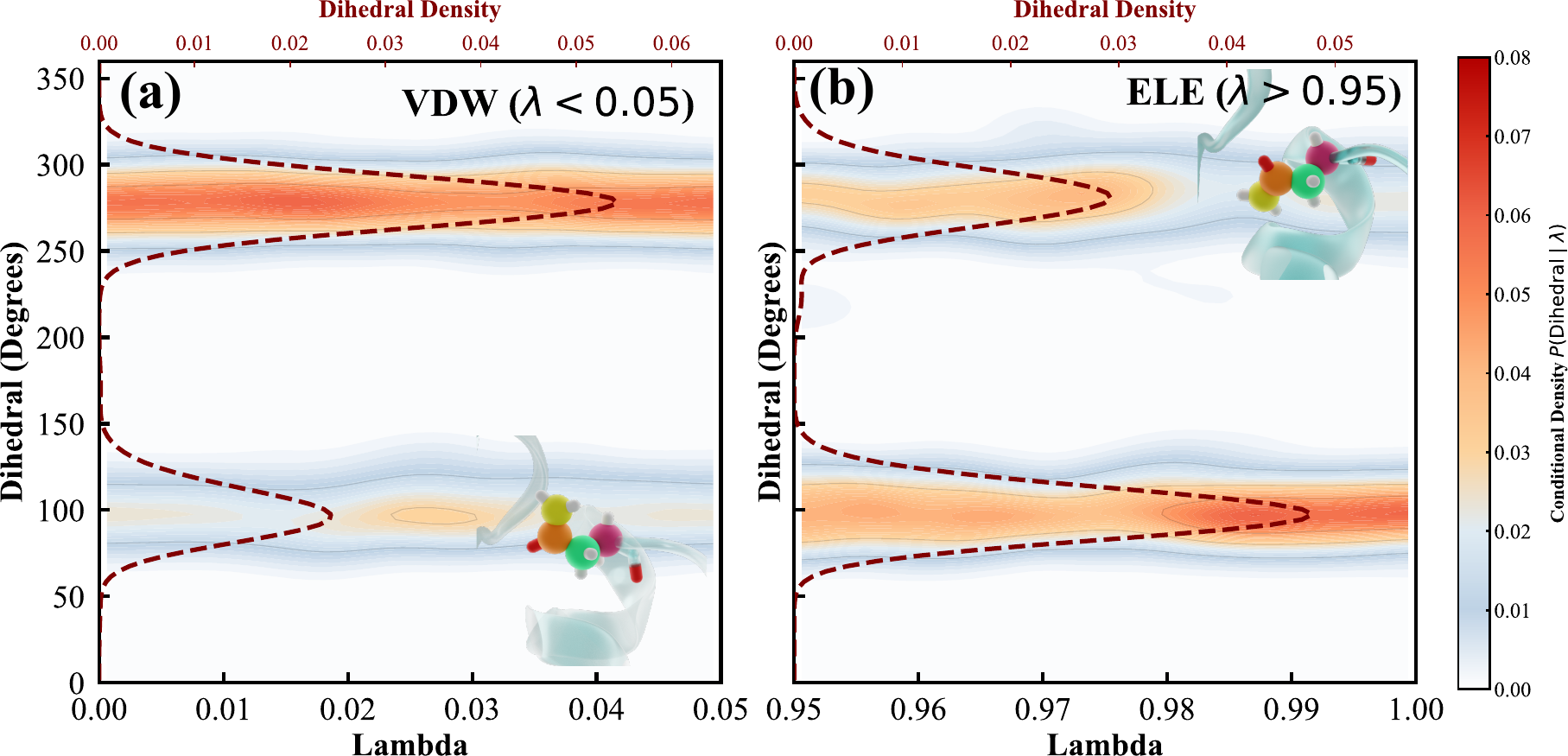}
 \caption{\textbf{Conditional side-chain rotamer landscapes of Asn140 in the ligand 5--BRD4 complex using the $\text{LAO+}$ framework}. Two-dimensional (2D) column-normalized conditional probability density landscapes, $P(\text{Dihedral} \mid \lambda)$, alongside matching vertically averaged 1D profiles (dashed maroon lines) are presented across refined endpoint sub-ranges for (a) the VDW endpoint ($\lambda < 0.05$) and (b) the ELE endpoint ($\lambda > 0.95$), both utilizing the OPES-biased conformational reservoir architecture. Insets in both panels visually map the representative molecular conformations of Asn140 corresponding to the distinct rotameric states cleanly isolated at the major conditional density bands, with the four atoms defining the core dihedral angle highlighted.}
    \label{fig:dihedral_sampling_BRD4}
\end{figure}

In our previous work~\cite{ansari2025lambda} on the BRD4 system using the standard Lambda-ABF-OPES method, we identified that the side-chain rotamers of Asn140 exhibit distinct conformational preferences in the apo versus the holo state. Specifically, the residue frequently adopts an orientation in the holo state that is energetically favorable upon ligand binding. To prevent sampling artifacts caused by these divergent preferences during continuous $\lambda$-switching, a positional restraint of 2 kcal mol$^{-1}$Å$^{-2}$ was previously required for the heavy atoms of Asn140 and its surrounding environment to maintain a consistent bound-like configuration. In this study, we demonstrate that the $\text{LAO+}$ framework can resolve this conformational bottleneck in BRD4 without the need for physical positional restraints. Instead of constraining the residue, we introduced a reservoir into both the VDW and ELE legs of the complex phase, specifically targeting the Asn140 $\chi_2$ dihedral angle. Using the OPES method, a  bias with a barrier of 20~kcal/mol was applied to this dihedral coordinate within the reservoir replicas.  The VDW reservoir explored the apo-like region ($\lambda \approx 0-0.2$), while the ELE reservoir was positioned at the intermediate alchemical stage ($\lambda \approx 0.5-0.8$). As illustrated in Fig.~\ref{fig:dihedral_sampling_BRD4}, this setup allows for the robust exploration of both rotameric states across the alchemical path. In the apo-like states ($\lambda \approx 0$), the probability distribution shows a clear preference for the 170$^\circ$ rotamer. Conversely, as the system approaches the holo state (ELE $\lambda > 0.95$), the distribution shifts, and the probability of the 75$^\circ$ rotamer becomes significantly higher. This behavior aligns with the native structural dynamics of the BRD4 binding pocket, where the side chain must reorient to accommodate the ligand. By allowing the BRD4--ligand~5 system to sample the correct conformational ensemble through alchemical exchange rather than fixed restraints, we achieved a calculated binding free energy of -8.60 $\pm$ 0.52 kcal/mol. This result is in excellent agreement with the experimental value of -8.8 $\pm$ 0.1 kcal/mol and notably consistent with the  $\text{LAO+}$ result obtained using restraints (-8.42 $\pm$ 0.44 kcal/mol). This confirms that $\text{LAO+}$ effectively captures the reorganization energy associated with side-chain reorientation while maintaining high thermodynamic accuracy without the need for manual structural constraints.

\subsection{Binding Mode Inter-conversion}

\begin{figure}[H]
    \centering
    \includegraphics[width=0.9\textwidth]{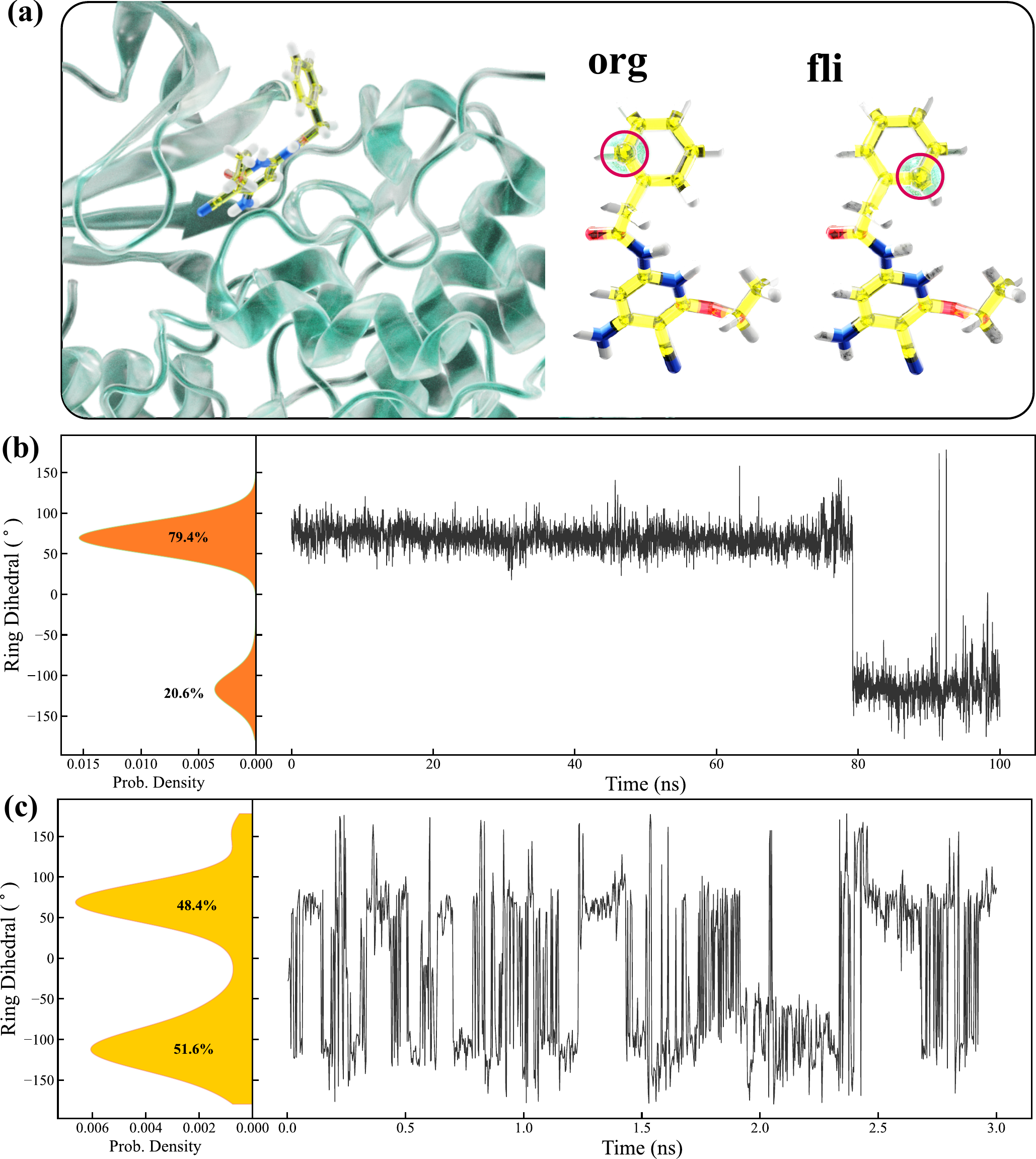}
    \caption{\textbf{Accelerated sampling of ligand ring-flip transitions in JNK1.} 
    \textbf{(a)} Left: Cartoon representation of Ligand 1 binding to the JNK1 pocket. Right: The two possible orientations of the terminal phenyl ring, labeled as original (\textit{org}) and flipped (\textit{fli}). Purple circles highlight the atoms defining the distinct orientations. 
    \textbf{(b)} Dihedral \del{+}trajectory and corresponding probability density from a 100~ns plain MD simulation, showing infrequent transitions and biased population (79.4\% vs 20.6\%). 
    \textbf{(c)} Dihedral trajectory and probability density from a 3~ns $\text{LAO+}$ simulation (replica-0), demonstrating rapid inter-conversion and nearly equal sampling of both states (48.4\% vs 51.6\%).
    }
    \label{fig:dihedral_org_fli}
\end{figure}

As a representative case study for inter-converting binding modes, we investigated the binding of Ligand 1 to JNK1 (see Fig.~\ref{fig:dihedral_org_fli}). In this system, the symmetry of the terminal phenyl ring implies that the two orientations (see Fig.~\ref{fig:dihedral_org_fli}a, right) should be energetically degenerate. However, a 100~ns plain MD simulation reveals that the system is severely kinetically trapped; only a single transition event is observed near 80~ns, leading to a heavily biased population distribution. To resolve this sampling limitation, we utilized the $\text{LAO+}$ framework. Here, we defined a dual-replica system consisting of a physical state (replica-0) and a specialized conformational reservoir (res). While replica-0 is set up for plain MD to sample the target ensemble, the reservoir is configured to facilitate barrier crossing by maintaining the alchemical coupling parameter between $0.5 < \lambda < 1.0$. Within this reservoir, we applied a nested biasing strategy: an ABF-OPES bias along the $\lambda$ coordinate to facilitate alchemical transitions, and an additional OPES bias directly on the phenyl ring dihedral to flatten the rotational free energy surface. These two replicas are connected via HREX, allowing the accelerated configurations from the reservoir to be swapped into the physical state. In this setup, we achieved a robust exchange acceptance ratio of approximately 45\%, ensuring a continuous flow of diverse conformations. As shown in Fig.~\ref{fig:dihedral_org_fli}c, the 'injection' of these biased configurations from the reservoir allows the populations of the original (\textit{org}) and flipped (\textit{fli}) orientations to rapidly inter-convert, converging to the expected 50--50\% distribution on the replica-0 trajectory in only 3~ns. 

\section{Discussion}

In this work, we introduced LAO+, a highly efficient multi-replica framework designed to overcome the sampling bottlenecks inherent in high-precision absolute binding free energy calculations. By replacing the global multi-walker architecture of the standard Lambda-ABF-OPES method with a partitioned HREX scheme and specialized conformational reservoirs, we have demonstrated a robust solution to the ``orthogonal barrier'' problem. Our results across five diverse protein--ligand systems show that $\text{LAO+}$ achieves a four-fold acceleration in convergence of the rate-limiting complex VDW leg compared to our previous $\text{LAO}$ scheme, while significantly reducing the computational footprint required for complex alchemical cycles.

The primary innovation of the $\text{LAO+}$ framework lies in its dual-layer approach to accelerating alchemical landscapes: first, by partitioning the global lambda range into localized windows, the framework targets non-linear phase transitions—such as the abrupt pocket hydration/dehydration 'water jump'—concentrating multiple replicas precisely where the kinetic barriers are highest rather than over-sampling trivial end-states. Second, it decouples the sampling of slow, orthogonal structural degrees of freedom—such as deep-seated side-chain rotamer transitions, and ligand ring-flips—from the primary alchemical coordinate. Through the use of these targeted reservoirs and the continuous injection of diverse configurations via HREX, $\text{LAO+}$ effectively eliminates Hamiltonian lagging, bridges artificial boundaries, and ensures that the most relevant physical states are sampled at every stage of the thermodynamic path. Because $\text{LAO+}$ acts on the sampling rather than on the energy model, it can in principle be combined with any force field, whether additive, polarizable or neural-network based; here it was demonstrated with AMOEBA.

Our validation studies, spanning targets such as BRD4, Trypsin, PWWP1, CHK1, P38, JNK1, and T4 Lysozyme, confirm that $\text{LAO+}$ recovers the relevant conformational state even in cases where standard MD and 1D alchemical methods remain kinetically trapped. The ability to efficiently sample degenerate binding modes and complex hydration landscapes marks a significant step forward in the accuracy and reliability of \textit{in silico} affinity predictions.

Complementary strategies pursue the same goal from a different direction. The Lucid double-decoupling method (LDDM)~\cite{bian2025formally}, for example, lowers the cost of fixed-charge ABFE by compensating the vanishing protein--ligand interaction with restraints so that their relative motion is minimized, but it explicitly leaves water exchange into and out of the binding site as an open problem. The focused $\lambda$-partitioning introduced here targets precisely this bottleneck, the reservoir mechanism addresses slow protein and ligand degrees of freedom, and the two ideas are in principle compatible.

Looking ahead, the modular nature of the $\text{LAO+}$ reservoir architecture provides a scalable platform for integrating Neural Network Force Fields and other emerging computational models into the drug discovery pipeline. By bridging the gap between atomistic precision and practical simulation timescales, $\text{LAO+}$ ensures that the thermodynamics of molecular recognition can be explored with unprecedented speed and accuracy, facilitating the rapid identification and optimization of novel therapeutic agents across a wide range of biological targets.


\section{Data Availability Statement}
The Tinker-HP and colvars input files used, and the structures
are available on GitHub at:  
\href{https://github.com/ansarinarjes/ABFE-L-ABF-OPES.git}{https://github.com/ansarinarjes/ABFE-L-ABF-OPES.git}

\section{Code availability}
The code used during the study is available on GitHub: 

\href{https://github.com/TinkerTools/tinker-hp}{https://github.com/TinkerTools/tinker-hp}

\href{https://github.com/TinkerTools/poltype2}{https://github.com/TinkerTools/poltype2}

\href{https://github.com/Colvars/colvars}{https://github.com/Colvars/colvars}

\section{Acknowledgment}

This work was supported by computing grants from the Grand Équipement de Calcul Intensif ($\text{GENCI}$), Institut du Développement et des Ressources en Informatique ($\text{IDRIS}$), and Centre Informatique de l’Enseignement Supérieur ($\text{CINES}$), France, under grant numbers  A0180716167 and A0200717469.

\bibliography{General,refs_v2_additions}
\end{document}